\documentclass{article}
\usepackage{arxiv}

\usepackage[utf8]{inputenc} % allow utf-8 input
\usepackage[T1]{fontenc}    % use 8-bit T1 fonts
\usepackage{hyperref}       % hyperlinks
\usepackage{url}            % simple URL typesetting
\usepackage{booktabs}       % professional-quality tables
\usepackage{amsfonts}       % blackboard math symbols
\usepackage{nicefrac}       % compact symbols for 1/2, etc.
\usepackage{microtype}      % microtypography
\usepackage{amssymb}
\usepackage{amsmath}

\usepackage{cleveref}       % smart cross-referencing
\usepackage{lipsum}         % Can be removed after putting your text content
\usepackage{graphicx}
\usepackage[numbers]{natbib}
\usepackage{doi}
\usepackage{graphicx}
\usepackage{booktabs}
\usepackage{float}
\usepackage{seqsplit}
\usepackage{changepage}
\usepackage{afterpage}
\usepackage{listings}
\usepackage{xcolor}
\usepackage{array}
\usepackage{tabularx} 
\usepackage{caption}
\usepackage{adjustbox}

\title{Large Language User Interfaces: Voice Interactive User Interfaces Powered by LLMs}

\author{
Syed Mekael Wasti\\
	Computer Science, Faculty of Science\\
	Ontario Tech University\\
	Ontario, Canada \\
	\texttt{syedmekael.wasti@ontariotechu.net} \\
	\And
    Ali Neshati \\
	Computer Science, Faculty of Science\\
	Ontario Tech University\\
	Ontario, Canada \\
	\texttt{ali.neshati@ontariotechu.ca} \\
	\AND
	Ken Q. Pu \\
	Computer Science, Faculty of Science\\
	Ontario Tech University\\
	Ontario, Canada \\
	\texttt{ken.pu@ontariotechu.ca} \\
}

\hypersetup{
pdftitle={},
pdfsubject={},
pdfauthor={},
pdfkeywords={},
}

\begin{document}

\maketitle

\begin{abstract}
The evolution of Large Language Models (LLMs) has showcased remarkable capacities for logical reasoning and natural language comprehension. These capabilities can be leveraged in solutions that semantically and textually model complex problems. In this paper, we present our efforts toward constructing a framework that can serve as an intermediary between a user and their user interface (UI), enabling dynamic and real-time interactions. We employ a system that stands upon textual semantic mappings of UI components, in the form of annotations. These mappings are stored, parsed, and scaled in a custom data structure, supplementary to an agent-based prompting backend engine. Employing textual semantic mappings allows each component to not only explain its role to the engine but also provide expectations. By comprehending the needs of both the user and the components, our LLM engine can classify the most appropriate application, extract relevant parameters, and subsequently execute precise predictions of the user's expected actions. Such an integration evolves static user interfaces into highly dynamic and adaptable solutions, introducing a new frontier of intelligent and responsive user experiences.

% We would like to encourage you to list your keywords within
% the abstract section using the \keywords{...} command.

\end{abstract}
\section{Introduction}
The modern world relies on and is driven by software. Embedded systems, command-line interfaces, web and desktop applications are present across systems all around the world. The predominant way in which we interact with software is through user interfaces. Their ease of use coupled with their intuitive nature has allowed for user interfaces to become an indispensable tool across contemporary software. These systems serve as a visually appealing packaging of function calls and event handlers, allowing complex event pipelines and data flows to be abstracted by buttons, text fields, menus, and so on. The catalyst for our research is grounded in the critical examination of current user interface systems, which, while omnipresent and essential, do not come without limitations. Traditional implementations of UIs often require the user to adhere to the methods and system constraints established by the developer and their design. This means that all UI components in an application have specific triggers and their invocation is limited to a strict sequence of user inputs within the applications' available features.\\
\par
The substantial advancements made in the space of large language models, especially over the past few years, have exhibited a true ``cognitive" potential. This potent ability has unveiled innumerable new opportunities to revolutionize the way our software systems are expected to operate. It became clear to us that an end-to-end architecture employing textual semantic modelling of contemporary user interfaces, as well as leveraging familiar user prompting behaviour, could be crafted. Within our research, we explore our vision and progress toward developing such a UI architectural paradigm, which benefits from the integration of a custom multimodal engine powered by LLMs and state-of-the-art transformer models. \\
\par
Our proposed framework aims to textually model the semantics of UI components, giving them the ability to communicate effectively with an LLM-driven agent environment. The automated generation of metadescriptors and annotations modelling each UI component allows us to form a data structure that streamlines agent-based parsing and entity extraction. Through this, we facilitate a synergistic interaction between the software and the user, enabling a more intuitive and user-centred experience, rather than requiring the user to conform to the software's predefined structures. Thus, our framework aids in abstracting monotonous UI interactions with prompting mechanisms that serve as ``cognitively aware", powering automated function calling and data flow pipelines. This translates to fully speech-based and intelligent control over visual user interfaces.\\
\par 
The benefits of our paradigm and its extensions are vast. Firstly, it would relieve the user of restricted, stringent and static rulesets when utilizing software. This would fundamentally change the landscape of how applications are developed, used and integrated across all sectors. It would allow for real-time intelligent and efficient control over complex tasks, enabling users to achieve substantially more, in significantly less time. They would be capable of multitasking through speech and executing complicated operations with real-time UI responses. Secondly, such a framework would further democratize software by drastically reducing learning curves, specifically for intricate software. A user would be able to provide their intended actions through speech and textual input and delegate heavy interactions to the framework, letting it reason with the UI to get the job done. Another major benefit of this framework is the streamlining of application development. User interfaces would be simplified substantially, hosting only the most necessary components. Furthermore, developers can generate annotations efficiently or automatically, allowing them to integrate functionality with far fewer input UI components.\\
\par
The paper is organized in the following manner. In \hyperref[sec:related_work]{section 2}, we showcase related works that introduce what has already been achieved in related domains, as well as where we fill gaps that are present. In \hyperref[sec:challenges]{section 3}, relevant challenges in the development of our framework and how they were overcome are thoroughly explained. \hyperref[sec:proposed_method]{Section 4} is where our framework is formally proposed in detail. \hyperref[sec:evaluation]{Section 5} discusses evaluations of the framework's accuracy and precision, along with showcasing comparisons between the integration of various LLM models. Finally, sections \hyperref[sec:future_directions]{6} and \hyperref[sec:conclusion]{7} discuss future directions, final thoughts, and conclude the paper.

\section{Related Work}
\label{sec:related_work}
The interest in applying large language models to user-centred domains is only natural, given that the greatest strength of these models is their semantic comprehension and reasoning abilities. Many research endeavours are focused on using LLMs in the context of human-computer interaction, crafting solutions for increased adaptation to user actions and behaviour. This includes training LLMs to better incorporate prompts from users through guided training pipelines which facilitate accurate execution of user commands \cite{ouyang2022training}, contextualization embeddings which grant more personalized responses \cite{ning2024userllm}, as well as implementing low-code interaction frameworks offering more intuitive control mechanisms over LLM instructing \cite{cai2023lowcode}. These works have pushed the capabilities of LLMs forward and opened up more powerful and efficient solutions for user-focused systems. Other efforts are being made in making LLMs increasingly flexible and versatile tools for human-computer interaction, pushing for ``journey" and context-aware interactions through few-shot learning methods \cite{brown2020language}, analysis of user-interest journeys, and desires \cite{christakopoulou2023large}. Training language models to teach themselves how to use tools in a self-supervised fashion and enabling zero-shot performance through demonstrations of API tools has also broadened the horizons of how LLMs can be integrated into existing pipelines \cite{schick2023toolformer}.\\
\par
Many works, similar to those mentioned, are geared towards direct interactions with LLMs, aiming to alleviate reliance on or bypass traditional UIs in favour of enhanced one-on-one sessions between the user and LLM systems. This can significantly improve efficiency and enable more complex operations. However, we find that there is a lack of focus on adapting and evolving contemporary user interfaces. While many system implementations would benefit from the dismissal of UIs, the focus on adding intelligence to their operations is crucial for the next generation of software. Users need to be able to visually observe their data as well as the UI's responses to their commands. They must be able to easily provide follow-up instructions with confidence that the framework will adapt in real time, equipping them with the ability to complete operations, simple or complex, with equal convenience. Eliminating learning curves or time-intensive tasks through intelligent speech control is invaluable.  \\
\par
These factors are what led us towards the research and creation of a two-component framework that revolutionizes both ends of the system. A paradigm that changes the way we develop applications, harnessing the power of LLMs to comprehend user prompts and permit back-and-forth agent-based reasoning with our dynamic annotation tree that semantically models the nature of all available UI components. This approach not only facilitates task efficiency, customization, and user-adaptive behaviour, but also simplifies user interface development and consumption, allowing applications to deliver valuable context, user control, and accessibility. We aim to leverage this architecture to make technology more approachable and usable for a broader audience.

\section{Challenges}
\label{sec:challenges}

The development of such a framework is not without its challenges. An effective fusion of LLMs and event-driven UIs is a complex task with many critical factors of concern. All classification, processing, parsing, and model inference must be achieved in real-time and the architecture must be capable of scaling and integrating unseen sub-applications in zero-to-few shots. Additionally, the framework's interpretation accuracy of ambiguous user inputs must be top-notch. The greatest challenges encountered during our research and implementation of this framework are discussed in this section.

\subsection{Integration Complexity}
The underlying challenge across the entire research endeavour was effectively fusing LLMs with event-driven UIs. Developing an effective framework capable of being scaled with increasing application counts, as well as fostering potential for implementation as an operating system was always a priority. A paradigm as dynamic as the software it intends to power was vital. Implementing a brute-force hard-coded solution would not serve a great purpose or carry the potential for scalability and impact required for next-generation software. This is why we took our time and carefully crafted each component of this framework with scaling in mind, leaving room for future advanced integrations, such as more powerful LLM model configurations and automated agent fine-tuning.

\subsection{Annotation Data Structure Selection \& Traversal}
Another critical concern was selecting a data structure to effectively represent the system, its applications, and all UI components. Through our research, a tree structure made for the most accurate representation of the overall system, whilst also allowing for efficient traversal and information retrieval, vital for real-time responses further on.\\

\begin{figure}

    \centering
    \includegraphics[width=1\linewidth]{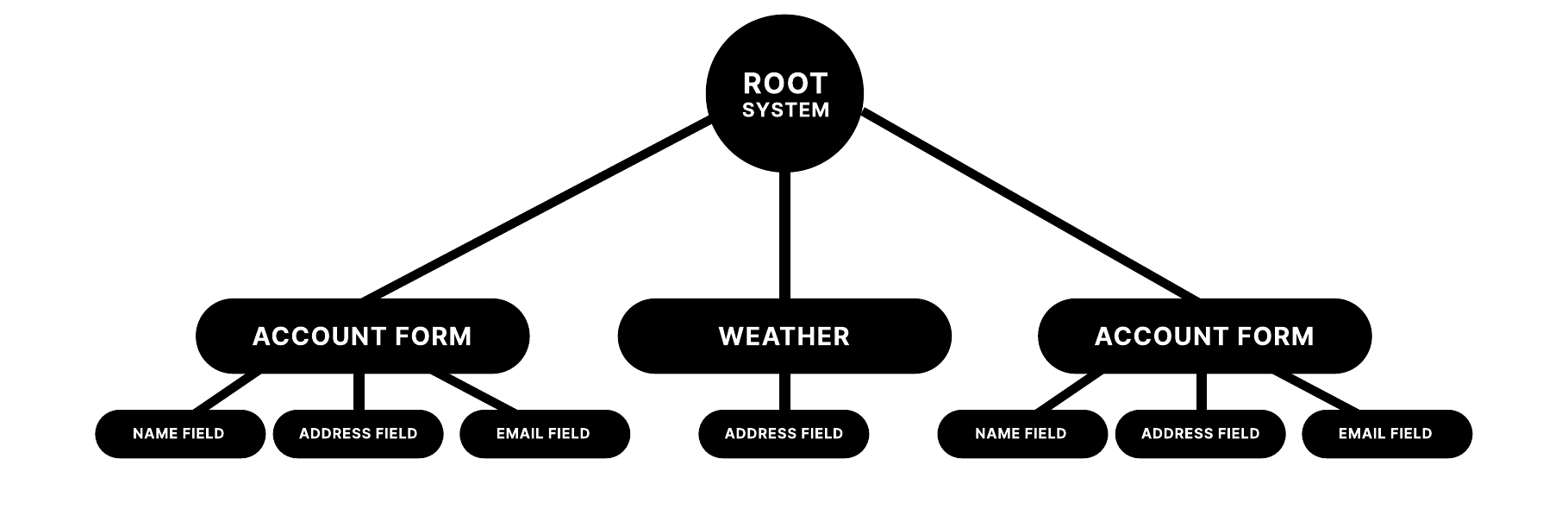}
    \caption{Visual representation of tree data structure used to store application meta descriptions}
    \label{fig:tree-data-structure}
\end{figure}

The specific implementation of this tree structure still poses room for iterative research concerning scalability, however, regardless of the implementation method, the need for top-down level-by-level comparison is where the tree structure shines. Rather than mapping an input to an application and its UI components through an unstructured or linear search, an organized level-by-level search eliminates redundant checks. Each user input is compared to a tree, which stores textual naming, descriptions, expectations, and examples of every application and its respective UI components. Our use of a top-down manner comparison between the tree structure and user input ensures that only children of relevant nodes are searched, preserving our engine's real-time processing goals. 

\begin{lstlisting}
`Extract the name, location, and email: "I am Brian O'Connor residing at Apartment 5A, 654 Peachtree Street in New Town and my email address is brian.oconnor@mailservice.net." || "Name": "Brian O'Connor","Address": "Apartment 5A, 654 Peachtree Street, New Town", "Email": "brian.oconnor@mailservice.net"',
`Extract the arithmetic expression: "If you have $50 and spend $25, how much money do you have left?" || "$50 - $25"',
`Extract the location: "Can you provide me with the current weather conditions in Zurich, Switzerland?" || "Zurich:B-CITY,Switzerland:B-COUNTRY"'

...
\end{lstlisting}

\subsection{Model Selection \& Training}
For our framework, an engine that is capable of high-level logical reasoning, comprehension, and general task capability was paramount. A multimodal pipeline was found to be most efficient in allowing for the balancing of inference times between models with parameter sizes appropriate for the task, as opposed to a single overkill model wasting processing time on smaller classification jobs. Tasks that could be accurately and fastest completed by smaller models were assigned models accordingly, while more complex tasks such as input comprehension, parameter extraction and parsing were given their own appropriate models. \\

\par

For each task, many models were researched with the goal to balance accuracy and speed. We found ROBERTA models \cite{liu2019roberta} fine-tuned on datasets such as SQUAD2 \cite{rajpurkar2018know} were efficient at entity extraction tasks in terms of speed, yet lacked precision when challenged with convoluted extractions. For such comprehension and parsing, we found that even powerful and large models such as Google's XL T5 \cite{raffel2023exploring} pre-trained to perform well with general multi-task capabilities simply cannot achieve the desired accuracy across our application library. The clear answer would be to employ larger and more powerful models, however, inference times and computational resources quickly become a concern.\\

Custom fine-tuning was key and opened a pathway for us to research how to achieve generalization well enough to output consistently formatted and precise results. \textit{See Table 1}. While training models on custom datasets relevant to each application in our library is effective, a scalable and more promising approach would be achieving few/zero-shot capabilities. We are exploring the most effective methods to accomplish this along with scaling our computational resources to open the door to models with larger parameter sizes and more rigorous training.
\begin{table*}[!h]
  \centering
  \normalsize
  % \begin{adjustwidth}{0cm}{}
  \renewcommand{\arraystretch}{3.0}

\caption{Model Benchmarks}

  \label{tab:table1}
    
    \begin{tabular}{c|c@{\extracolsep{20pt}}c@{\extracolsep{20pt}}c@{\extracolsep{20pt}}c}
    \toprule
    
    & \parbox[c]{1.667cm}{\centering T5\\ \scriptsize Custom Fine-Tune} & \parbox[c]{1.667cm}{\centering T5 \scriptsize \\Google MTL}& \parbox[c]{1.667cm}{\centering BERT \\\scriptsize SQuAD2}& \parbox[c]{1.667cm}{\centering ELECTRA \scriptsize SQuAD2}\\
    \midrule
    Weather & 95\% & 36\%& \textbf{96.8}\% & 76\%\\
    Account Form & \textbf{80}\% & 0\%& 78\% & 48\%\\
    Simple Calculator & \textbf{100}\% & 1\%& 99\% & 16\%\\
    Advanced Calculator & \textbf{81}\% & 1\%& 5\% & 1\%\\
    
    \bottomrule
  \end{tabular}

\end{table*}
\subsection{Training Data Generation \& Wrangling}
Creating custom datasets with realistic, convoluted and vague prompts reflective of real-life scenarios was challenging. We found no available datasets with the nature of prompts required to be effective for our training. Custom datasets were the best option and were a significant task. This is where generation with top-of-the-line LLMs such as ChatGPT-4 served immense value. Many competing models were tested, however, the adherence to prompt engineering, logical understanding of the task, as well as the depth of versatility GPT-4 \cite{openai2023gpt4} was able to provide, were unmatched in our findings. OpenAI's API was used for mass data generation and any formatting inconsistencies were fixed by hand. These custom datasets were used to fine-tune LLM models, such as varying sizes of Google T5 on various tasks. \\\\

\section{Proposed LM-UI  Framework Architecture \& Design}
\label{sec:proposed_method}

With the aforementioned motivations and challenges, we present the Language Model-User Interface framework. Through the fusion of an annotated front-end component system and a multimodal extraction engine specialized in classification and logical entity recognition, we are able to lay the groundwork for the next generation of user experiences. The framework employs an end-to-end custom architecture that hinges upon the integration of a specialized tree data structure which effectively manages annotations and metadescriptions of UI components. The creation, indexing, and merging of this data structure is what allows us to employ LLM agents capable of real-time interpretation and prediction regarding a user's intended actions. Through this pipeline, user interfaces can be significantly simplified, whilst becoming increasingly dynamic and efficient.

\begin{figure}[h]
    \centering
    \includegraphics[width=0.65\linewidth]{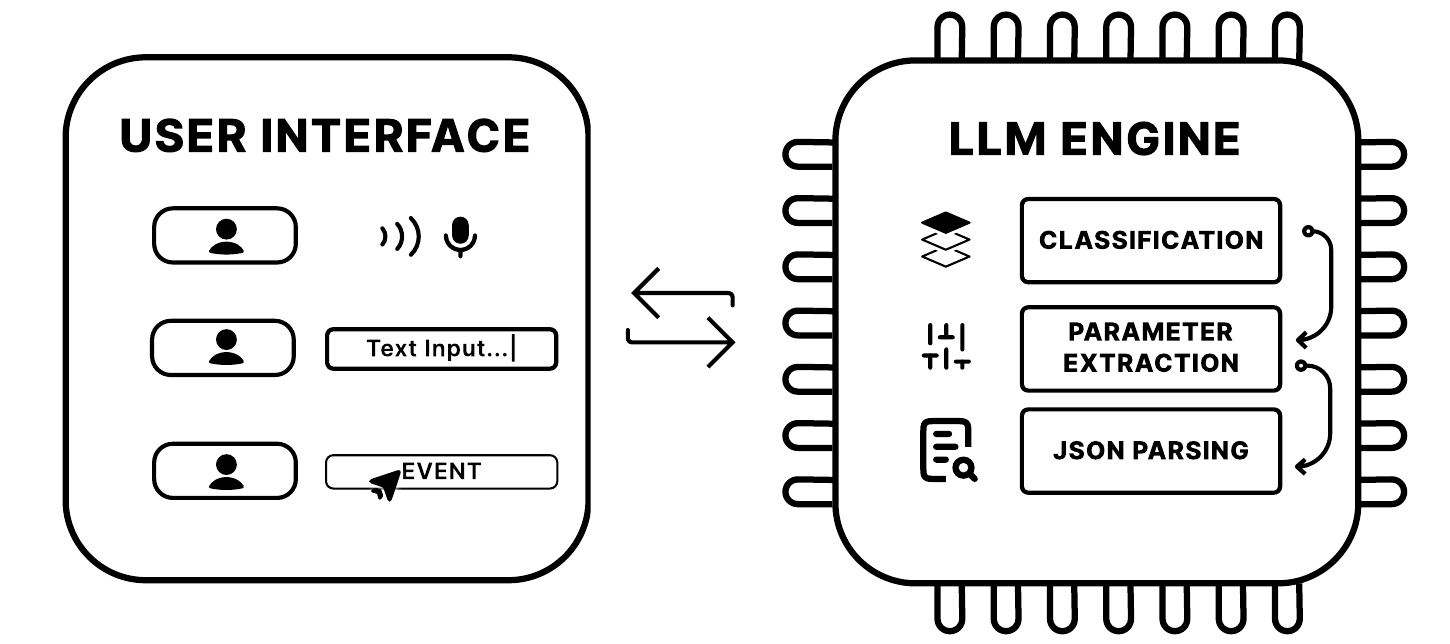}
    
    \caption{Two-Component Framework}
\end{figure}

\subsection{Describing Applications as Events \& States}
The first aspect of our architecture asks that the applications in our framework be modelled upon the concept of events and states. Each UI element in an application has a state for all its configurable parameters. Within each application's root component, its states are observed for any changes. When states are updated in response to a user event, the component is immediately re-rendered to reflect the new states. When the master JSON object holding a current application indicator along with the states of its respective parameters is updated, the new states are dispatched to the central store. Each interaction with the UI components of an application is an event that is reflected by the Redux states and store.

\subsubsection{Central Store \& Reducers}
When a user’s textual prompt is received, it is sent to the LLM engine for processing. Once the engine returns the parsed result to the root front-end component, it triggers action events. These events dispatch the new states to reducers which are functions that take the previous state and an action as arguments and calculate a new state. They are responsible for handling transitions from one state to another within the application, ensuring state changes are predictable and transparent. The Redux central store is then updated to reflect the new state. When this store is updated, useSelector hooks and connect functions detect the state change and immediately update the UI. 
\begin{figure}
    \centering
    \includegraphics[width=0.85\linewidth]{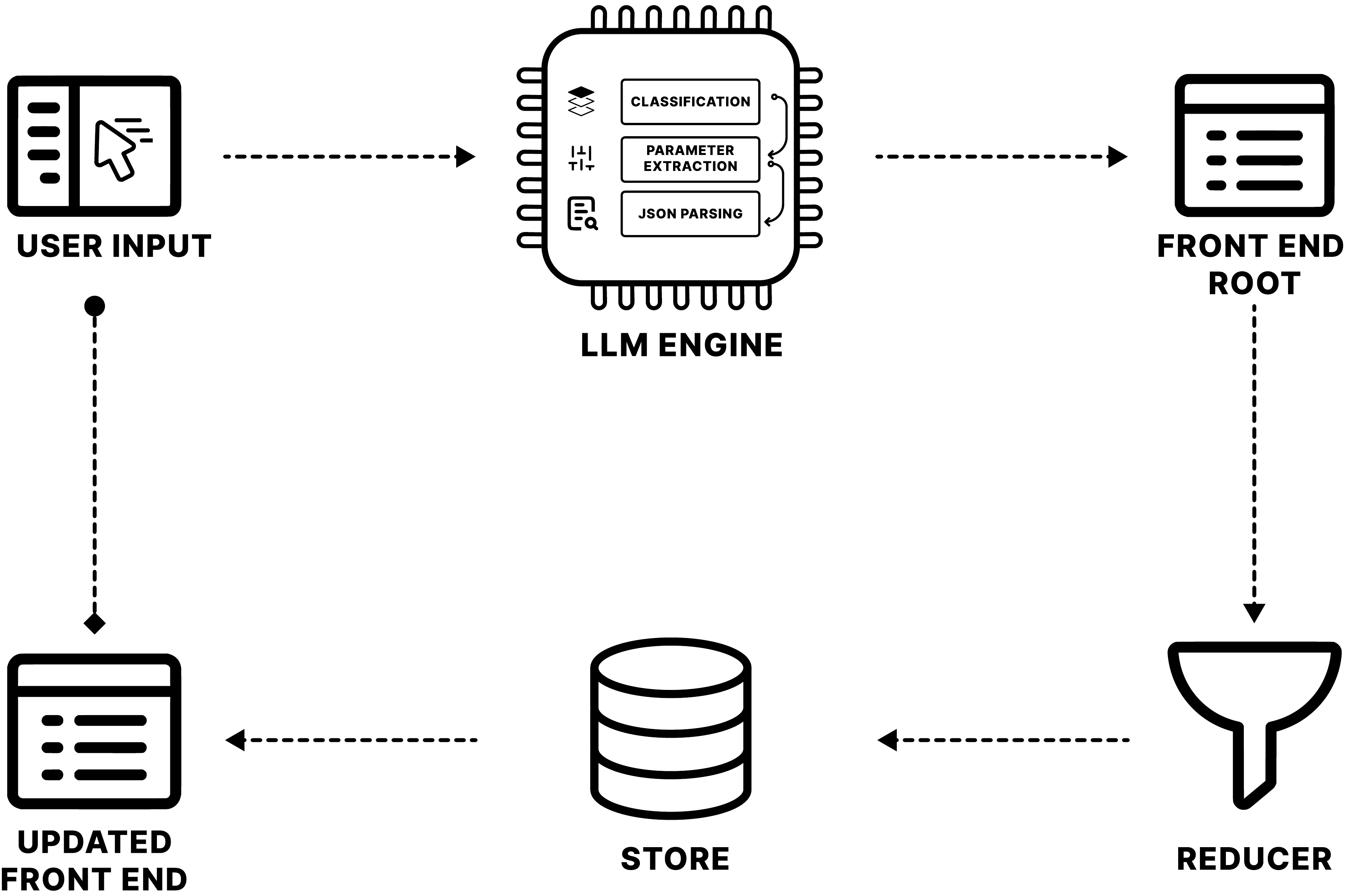}
    \caption{Application State Flow}
    \label{fig:Redux Store Flow}
\end{figure}
\par
In the context of our framework, when an action is dispatched to the Redux store, it is forwarded to the corresponding reducer. The reducer, being a pure function, takes the current state and the dispatched action as its arguments. It then calculates and returns a new state without modifying the old state. The Redux store updates its internal state with this new state returned by the reducer. This updated state then propagates through the application, automatically triggering updates in the UI components that are dependent on this part of the state. This process facilitates a smooth, unidirectional data flow, ensuring that UI components can remain stateless, which promotes their reusability and testability.

\subsubsection{Case Study: Application/Task Library}
% Account Form, Weather, Account Form\\

The implementation of this framework serves as a type of “operating system” that functions over a library of applications. Thus far, we have three simple applications in the library: An account sign-up form, a live weather application, and a calculator. These applications help showcase and validate the core functionality and flexibility of our architecture. It is easy to see how this can be scaled up to include various essential applications that are commonly found on smartphones, tablets, smartwatches, etc. Once we can establish and validate the framework's potential and ability, more complex and intricate UI applications can be tackled with the appropriate merging and implementation schemes.

\subsection{Modelling User Interface Components}
Our approach to fusing LLMs with UI stands firmly upon the modelling of UI components as adaptive entities capable of dynamically responding to natural language inputs from the user. At the core of this model is a re-imagined method to define UI elements such as buttons, text fields, navigation links, etc. These components are thoroughly described and textually modelled in our annotation tree data structure. As a result, their integration and flexibility are streamlined in an event-driven environment where behaviours and states are modifiable based on user interactions conveyed through natural language prompts.
\begin{figure}
    \centering
    \includegraphics[width=0.85\linewidth]{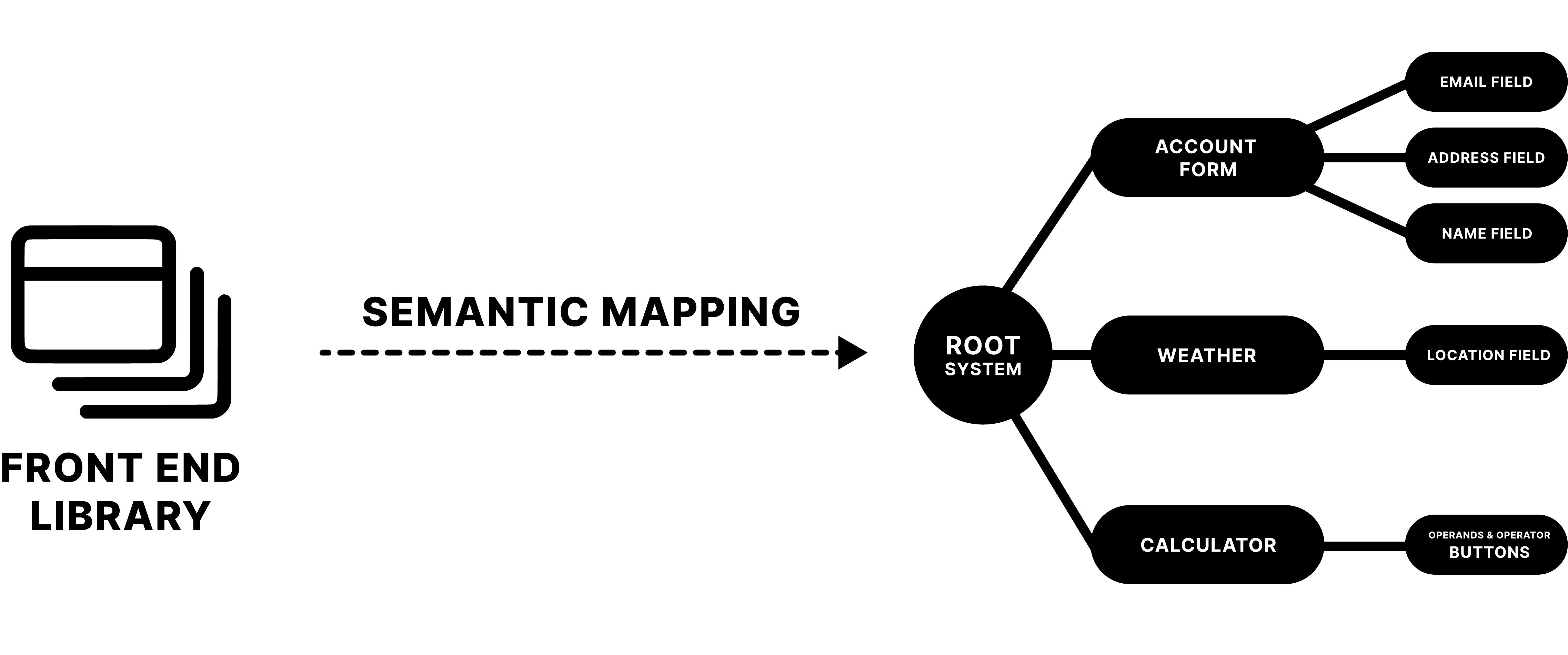}
    \caption{Each application is semantically mapped into the tree structure. Each node holds detailed annotations}
    \label{fig:Semantic Mapping}
\end{figure}
\subsubsection{Annotation Tree \& Meta Descriptions}%
The data structure storing meta descriptions of each UI component in an application is the key principle behind the proposed framework. Our requirement of this textual modelling is the crucial link bridging the gap between contemporary user interfaces and the emergence of highly advanced LLMs. These annotations provide detailed specifications of how each UI component fits conceptually within the structure of the application, what its configurable parameters are, and examples of potential use cases. This information primes applications for communication with the back-end engine's agents further down the pipeline.\\\\\\
Let us denote our application library system as \(S\), where \(S\) contains all elements representing applications and their components:
\begin{equation*}
S = \{A,W,C\}
\end{equation*}
where A, W, and C denote the sets for the Account Form, Weather, and Calculator applications from our application library, respectively. Let each nested application set be denoted as:
{\setlength{\jot}{10pt}
\begin{align*}
& A = \{a_1, a_2, a_3...\} \textit{ for Account Form components}\\
& W = \{w_1, w_2, w_3...\} \textit{ for Weather components}\\
& C = \{c_1, c_2, c_3...\} \textit{ for Calculator components}
\end{align*}
}where each contained element represents a unique node modelling a UI component. Each node stores annotations describing the component's role in the application, along with the descriptors of what parameters it manipulates. See \textit{Listing 1.2} 
\begin{lstlisting}
a1: { 
      "AppName": "Weather",
      "Component Name": "Location Field"
      "Description": "A sub-node of Weather. Represents the point where location or query details are input for weather-related inquiries.
    "Parameters": { "City": "What is the location?"}
    },
...}
\end{lstlisting}

\subsection{Parsing \& Mapping User Input}

The user provides input describing their target intention through the system's textual prompt input field or through a speech button which triggers a recording of the user's spoken prompt. Once received and converted to text through speech recognition models, if speech is used, a prompt is sent to the back end where it is processed by an agent-based prompting mechanism. The processing pipeline carries out appropriate \textbf{encoding}, \textbf{tokenization}, and finally \textbf{mapping via classification}.
\\\\
\textbf{Tokenization}

\begin{enumerate}

    \item An input sentence is split into tokens, adhering to an agent's specific LLM encoding requirements. We will cover BERT and its use of WordPiece tokenization for this example\\ 
    \item Special tokens like `[CLS]` (at the beginning) and `[SEP]` (at the end) are added\\
    \item The tokenization can be represented as a function that converts a sentence into a sequence of tokens:
\begin{align*}
\textit{Tokenize}(\textit{``Input Sentence''}) = [&\textit{``[CLS]''},\textit{``Token}_1\text{''}, \textit{``Token}_2\text{''},...,&\textit{``Token}_N\textit{''}, \textit{``[SEP]''}]\\
\end{align*}

\end{enumerate}
% \textbf{Token to ID Conversion}
\textbf{Encoding}

\begin{enumerate}

\item Each token is then mapped to an ID using BERT's vocabulary. This can be represented as
     \[ \text{ID}_i = \text{Vocab}(\text{``Token}_i\text{")}, \text{ for } i = 1, 2, ..., N \]
where `Vocab` is a function that returns the ID of a token in the BERT vocabulary.\\

\item The total embedding for a token, \(E_{token}\), includes word, segment, and positional embeddings.\\
\end{enumerate}

\subsubsection{Mapping via Classification}

To classify user input, we encode the input and node representations:
\begin{itemize}
    \item For user input \(u_i: v_{u_i} = encode(tokenize(u_i))\)
    \item For a node \(n_j\) representing an application or component: \(v_{n_j} = encode(tokenize(n_j))\)
\end{itemize}
Let U denote the user's encoded input:
\begin{equation*}
U = \{u_1,u_2,u_3...\}
\end{equation*}

The system then maps user input to system nodes using an interpretation function that performs classification, \(f: U \longrightarrow S\), where:\\

\(f(U_i) = n_k\), where \(n_k\) is chosen such that: 

\begin{equation*}
    n_k = argmax(cosine\_similarity(\vec{v}_{u_i}, \vec{v}_{n_j}))
\end{equation*}\\
This process classifies the user's input to the most relevant application component in the annotation tree. To handle complex inputs that involve specific actions or components, we extend the function \(f\) to further map inputs to nested sets (components) within \textit{A}, \textit{W}, or \textit{C}:\\
\begin{equation*}
f(u_i) = a_j \in A \textit{ or } f(u_i) = w_j \in W \textit{ or }  f(u_i) = c_j \in C
\end{equation*}\\
This implies that the function not only identifies the relevant application but also the specific action or component within that application.
\subsubsection{Entity Extraction and Parameterization via LLM Inference}
At this level, the custom fine-tuned LLM agents with specialized ability such as Google's T5 model trained on specific tasks from our library in a teacher-student environment, enable accurate parsing of complex user inputs into structured data that the system can act upon. This process is crucial for interpreting the specific details contained within the user’s input and mapping them to actionable UI component parameters in our system. 
\begin{figure}
    \centering
    \includegraphics[width=1\linewidth]{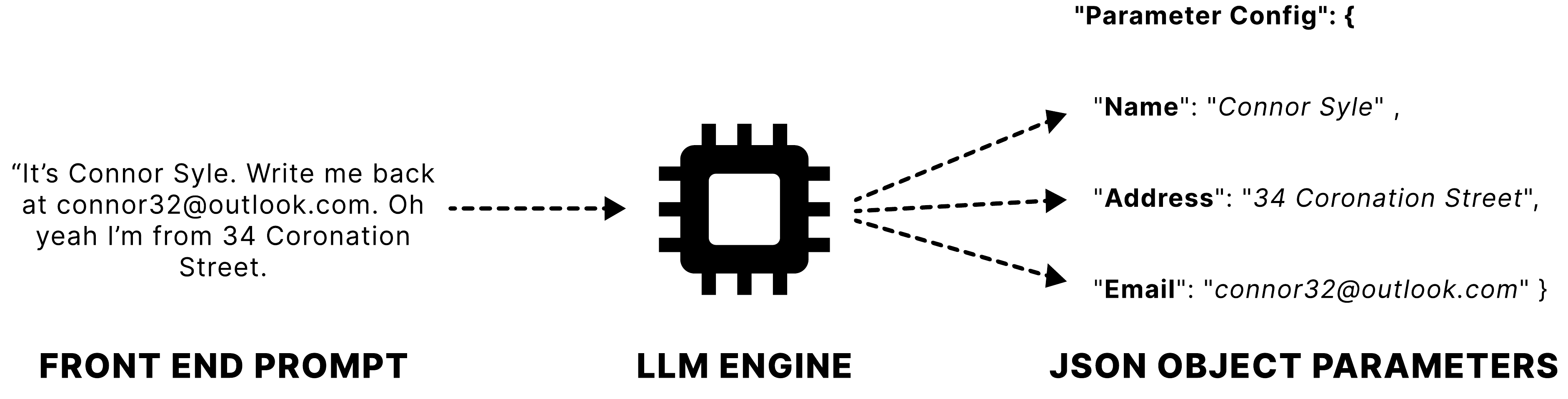}
    \caption{Entity Extraction \& Paramaterization}
    \label{fig:Entity Extraction & Paramaterization}
\end{figure}

For inputs that require extracting parameters or entities, we define a transformation \(g\): \(U\)→\(P\), where \textit{P} is a set of parameters extracted from the inputs. This can be used alongside \(f\) to not only map an input to an application component but also extract the necessary information to perform the action:
\begin{equation*}
\textit{g}(\textit{u}_\textit{i})=\{\textit{p}1,\textit{p}2,…\}    
\end{equation*}
where \(u_i\) is the encoded user input, and \(\{p_1,p_2,...\}\) are the extracted parameters. These parameters are then used to dynamically adapt the system's response to the user's intent. This extraction algorithm functions as a decision-making tool with the use of natural language processing techniques. Its objective is to analyze textual input, categorize it based on predefined criteria, and generate a JSON parameter extraction object to be relayed to the front end and dispatched to the Redux central store.

\subsubsection{Action Execution Based on Mapping}

Finally, we use the outcomes of \textit{f} and \(g\) to create a final JSON object and execute actions within the system. This involves sending the final JSON object to the front end where the Redux central store invokes specific methods associated with the identified components, passing the extracted parameters and dynamically updating the UI. Let the following represent the introduction of relevant methods:

\begin{equation*}
execute(f(u_i), g(u_i)) 
\end{equation*}

\subsubsection{Example:}

Assume a user input \(u_i\)= ``I have a flight to France, what's it like out there?". The interpretation function \textit{f} maps this input to the weather application's ``location" field component \(w_{location}\in W\), and \(g\) extracts ``France" as a parameter. The system then executes the location-based action for ``France" based on this mapping and parameter extraction.\\
\par
The defined detailed approach allows the system to dynamically interpret user inputs by understanding their intent and mapping them to the most relevant application and components in the application library system. It ensures that user inputs are handled efficiently, navigating through the hierarchical structure of the application library based on semantic similarity, thereby improving user experience through accurate identification and interpretation of a user's needs.

\subsection{\textbf{Operational Workflow}}

\begin{enumerate}
  \item \textbf{Input Analysis} The algorithm initially analyzes the input to determine its nature and relevance to each UI component child node under the classified application. This categorization is crucial in deciding the subsequent processing path.\\
  \begin{enumerate}
      \item \textbf{Top-Down, First-Match Traversal} The system starts the mapping at the root of the application tree. It uses our defined interpretation function \textit{f} to select the child node with the highest cosine similarity to the user input. The process repeats until a leaf node is reached, moving to children nodes only after the parent node is classified.\\
      \item     \textbf{Early Stopping for Efficiency \& Error Mitigation} A threshold is in place for cosine similarity. If the similarity between the user input and the best-matching node at any level is below a predefined threshold, the search is halted, and the system can proceed with action execution based on the current best match, or prompt the user to try again. This optimizes the search process, ensures responsiveness and mitigates errors that can result from forced mapping.\\
  \end{enumerate}

  \item \textbf{Conditional Logic} Based on the analysis, the algorithm employs conditional logic to deploy appropriate agents in the extraction mechanism. Parent nodes of classified applications are checked, to guarantee the most suitable model is utilized. This selection process is pivotal to the algorithm's adaptability and accuracy in parameter extraction.\\

  \item \textbf{Response Generation} Depending on the categorization:\\
    \begin{itemize}
      \item If the input falls into a specific category (e.g. calculations), the algorithm engages a dedicated module designed for that category, employing specialized deep-learning models, such as our custom fine-tuned Google's T5\\
      \item For other types of input, the algorithm utilizes a different set of models tailored for general language understanding and response formulation\\
      \item This delegation of tasks makes it so that we use the most appropriate model for the task, employing larger more capable models for more demanding tasks such as complex arithmetic extraction, while making use of suitable smaller models for simplistic tasks. This optimization ensures valuable inference time from larger models is not wasted on tasks that don't require it, saving resources along with crucial response time.\\
    \end{itemize}

  \item \textbf{JSON Output Formatting} Finally, the parameters extracted by the model are formatted into a structured JSON object, optimized and suitable for further processing or for display to the end-user once sent back to the client. \textit{See Listing 1.3}.
\end{enumerate}
\begin{lstlisting}[caption={Extracted parameters parsed into a JSON object}, label={lst:json_output}]
{"CurrentApp":"AccountForm","Config":{"Name": "Connor Syle", "Address": "34 Coronation Street", "Email": "connor32@outlook.com"}}
\end{lstlisting}
\par\vspace{\baselineskip}
This pipeline employed by our LLM engine allows for complex comprehension of textual input from a user, by accurately classifying and extracting parameters that the target application requires. Our engine is able to streamline the connection between the user and the software by accounting for the intentions and requirements of both, utilizing its logical reasoning to complete the task as it sees fit. This is a stepping stone in the right direction for the next generation of software.

\section{Evaluation}
\label{sec:evaluation}

This section intends to evaluate and validate the abilities of several large language models tasked with serving as the crux of our research. Accuracy in parsed inputs is the driving factor behind this project functioning well. The engine must correctly classify which task the user's prompt corresponds to, extract, parse, and output the expected parameters in the required JSON format. 

\begin{table*}[h]
\centering
\renewcommand{\arraystretch}{1.5} % Adjust the row height
% \small
% \footnotesize
% \scriptsize
\begin{adjustwidth}{0cm}{}
\begin{tabular}{|m{4.5cm}|m{4.8cm}|m{4.8cm}|m{0.45cm}|}
\hline
\multicolumn{4}{|c|}{\textbf{GPT Baseline VS Custom Engine Prompt Evaluation}} \\
\hline
% \textbf{Truth Intention} & \textbf{GPT-4 Generated Inputs} & \textbf{BERT SQuAD2 Parser Output}& \textbf{Eval} \\
% \textbf{Truth Intention} & \textbf{GPT-4 Generated Inputs} & \textbf{Multimodal Parser Output}& \textbf{Eval} \\
\multicolumn{1}{|c|}{\textbf{GPT-4 Generated Inputs}} & \multicolumn{1}{c|}{\textbf{Truth Intention}} & \multicolumn{1}{c|}{\textbf{Parser Output}} & \multicolumn{1}{c|}{\textbf{Eval}} \\
\hline
I'm registered under the name Alex J. Turner, but everyone sends their regards to my place at 768 Rolling Rock Street, and for a quicker response, they hit me up at alex.turns@rocknmail.com. & \seqsplit{\{``CurrentApp":``AccountForm",``Config":\{``Name": ``Alex J. Turner", ``Address": "768 Rolling Rock Street", ``Email": ``alex.turns@rocknmail.com"\}\}} & \seqsplit{\{``CurrentApp":``AccountForm",``Config":\{``Name": ``Alex J. Turner", ``Address": "768 Rolling Rock Street", ``Email": ``alex.turns@rocknmail.com"\}\}} & \checkmark \\
\hline
Could you tell me if I need an umbrella for my walk today around the canals of Amsterdam, Netherlands? & \seqsplit{\{``CurrentApp":``Weather",``Config":\{``City": ``Amsterdam, Netherlands"\}\}} & \seqsplit{\{``CurrentApp":``Weather",``Config":\{``City": ``Amsterdam, Netherlands"\}\}} & \checkmark \\
\hline
Just finished a book set in Venice, Italy, and it got me wondering about the weather there today. Is it as sunny as in the story? & \seqsplit{\{``CurrentApp":``Weather",``Config":\{``City": ``Venice, Italy"\}\}} &  \seqsplit{\{``CurrentApp":``Weather",``Config":\{``City": ``Venice, Italy"\}\}} & \checkmark \\
\hline
My friend in Sydney mentioned they were having quite the heatwave. I'd love to compare that to the current temperatures where the Great Pyramids stand. Could you give me a quick overview of the weather there? & \seqsplit{\{``CurrentApp":``Weather",``Config":\{``City": ``Egypt"\}\}} &  \seqsplit{\{``CurrentApp":``Weather",``Config":\{``City": ``the Great Pyramids stand"\}\}} & $\times$ \\
\hline
I've got 24 cupcakes, and I need to divide them evenly among my 6 friends. How many does each person get? & \seqsplit{\{``CurrentApp":``Calculator",``Config":\{``promptSequence": "24 / 6"\}\}} & \seqsplit{\{``CurrentApp":``Calculator",``Config":\{``promptSequence": "24/6"\}\}} & $\checkmark$ \\
\hline
\end{tabular}
\captionsetup{skip=10pt}
\caption{Comparison of GPT-4 generated inputs with our custom multimodal LLM engine consisting of custom fine-tuned Google T5 \& BERT-Large-Uncased-Whole-Word-Masking-Squad2 outputs. The model performs with extreme efficacy on descriptive and lengthy prompts.}
\end{adjustwidth}
\end{table*}
\vspace{-0.5mm}

\subsection{Required Tasks}
Classification is accomplished by encoding descriptive nodes for each task along with the input node using an all-MiniLM sentence-transformer model and then selecting the node with the greatest cosine similarity. This classification task is extremely reliable and the chance of failure is observed to be negligible. The critical task to evaluate is the parameter extraction once the input's corresponding application has been established. Extensive experimentation and trials were conducted with plentiful transformer models from the Hugging Face hub. Models specialized in Question Answering as well as Text2Text Generation were found to perform extremely well.

\subsection{Model Selection}
For this evaluation, we pit the most effective models found during initial testing, with available resources for training, hosting, deployment and inference in mind—the T5 Text2Text Generator models along with BERT and ELECTRA. The BERT and ELECTRA models were already fine-tuned on the SQuAD 2.0 dataset, which is a popular dataset used in fine-tuning models for logical reasoning and question-answering. On the other hand, the T5 model base and multi-task, trained by Google, were also chosen due to their flexibility, scalability, and efficacy in abstraction and multi-task situations.\\\\

\begin{figure}
    \centering
    \includegraphics[width=1\linewidth]{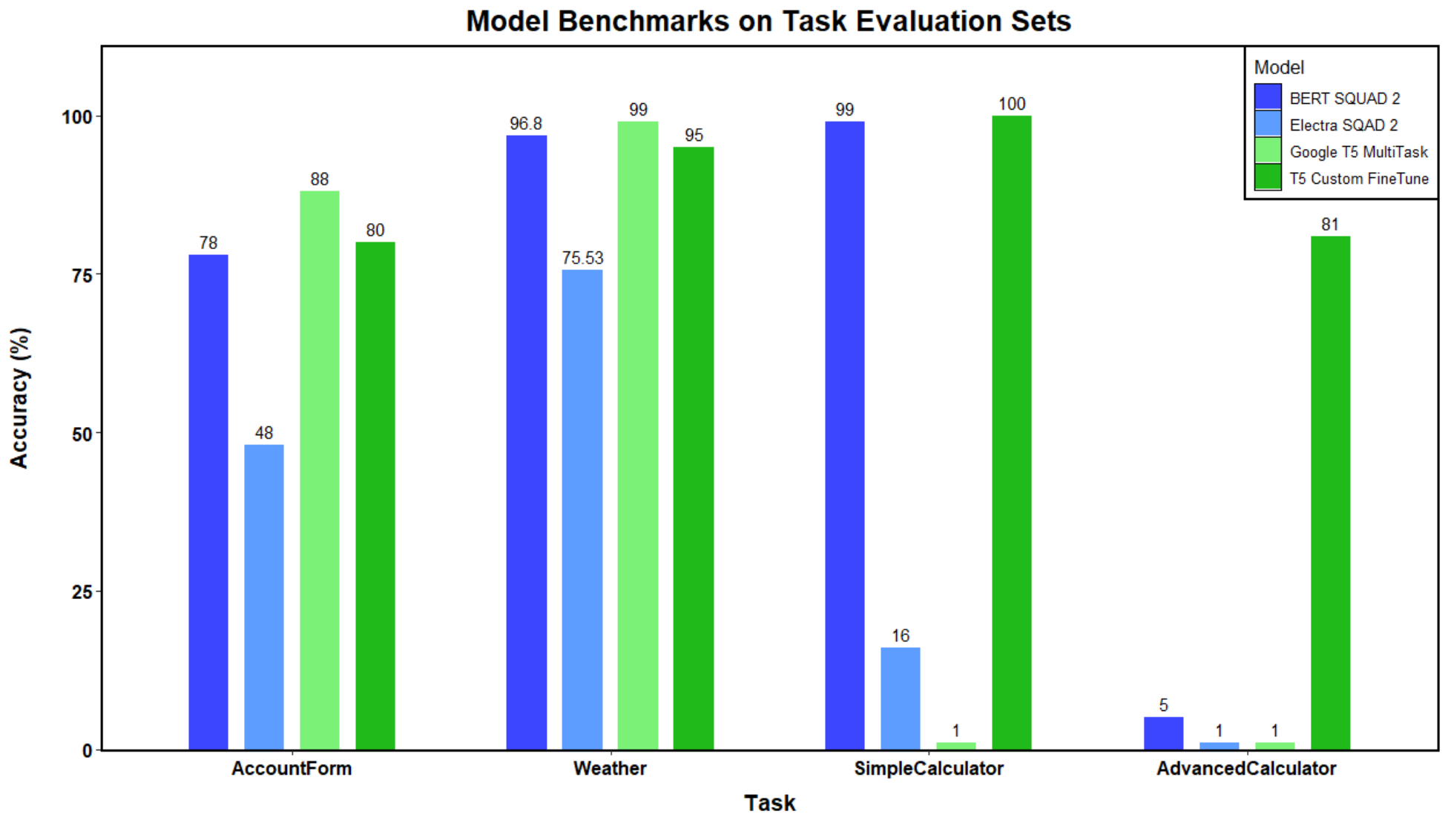}

    \caption{Accuracy comparison between models engineered for our application's tasks. Each model was trained on logical and information extraction tasks, and further carefully prompt engineered to achieve the best performance possible for each task.}
    % \caption{Model Evaluation}
    % \label{fig:Model Evaluation}
\end{figure}
\vspace{-7.5mm}

\subsection{Observation}
We can see from the benchmarks in \textit{Figure 6} that the T5 model trained on specialized custom datasets relevant to specific application tasks does exceptionally well across all tasks. It performs the best in three out of four tasks, with the last one being a close tie. The BERT model also performs very well but struggles with the logical capabilities required for the advanced calculator prompts. Google's multitask-trained model falls extremely short in more logically involved tasks, while the ELECTRA model does poorly overall. Further training of the T5 model with more diverse and better quality data would allow it to be the most versatile and powerful model for our framework.

\section{Future Directions}
\label{sec:future_directions}

To improve this framework and scale its potential, continuous iterations on both front-end and back-end engine components need to be considered. 

\subsubsection{Automation of Annotated UI Paradigm}
An automated structure for generating classical UI components through textual descriptions would allow for a unified annotated system. If a developer's textual prompts can generate UI components, they can be parsed to serve as annotations for the components as well. This would immediately make the UI compatible with the LMM engine's requirements and would transform the way full-stack applications are developed. 
\subsubsection{LLM Engine Embellishments}
Exploring configurations regarding transformer models that classify, extract, and parse information in the LLM engine will improve accuracy and inference speed. The choice between delegating tasks to different models versus using a singular multi-task model for all tasks is open to be explored. 

\subsubsection{AutoGen}
Expanding on the back end embellishments, making use of Microsoft's AutoGen \cite{wu2023autogen}, Mistral's Mixtral 8x7B mixture of experts model \cite{jiang2024mixtral}, or Ollama's model hot-swapping to allow for multi-model agent setups may be the ultimate solution for delegating a task to the appropriate model. AutoGen's optimized mixture of experts system would pose many benefits over a manual swapping system, especially with the capability for agents to communicate with one another. If an AutoGen setup can delegate a task to the best-suited LLM amongst each other, accuracy can be maximized across all tasks and inference times can be further optimized.

\subsubsection{Concurrent Processing}
Further scaling would call for large-scale concurrent processing of incoming user requests. Whilst our back-end server is currently multi-threaded and can handle many inputs simultaneously, further scaling to incorporate asynchronous processing with expanded capabilities, along with more robust load balancing and distributed computing systems, would bulletproof our framework's servers.

\section{Conclusion}
\label{sec:conclusion}

Our proposed LM-UI framework re-imagines the development process of contemporary user interface systems. We employ a system that automates the creation of textual semantic UI component annotations, which are supplementary to an agent-based prompting back end. This hybrid approach harnesses the power of  LLMs to map generic, vague, and convoluted user textual input to their relevant components. We can further parse the input through logical entity extraction processes, creating a JSON object capable of dynamically adapting the UI. Our evaluations showcase the accuracy and precision of this approach, allowing us to take a large step closer to bringing what was once a science fiction dream, to reality. \\
\par
Furthermore, the proposed framework is just the beginning. The extensions and features we can implement atop this core framework increase the benefits substantially. From streamlining complex workflows in industries such as finance and healthcare to revolutionizing highly productive multitasking environments in modern workplaces, the implications are vast and worthy of exploration. We aim to continue our research efforts, to inspire and propagate a new wave of research and development in this direction. Our objective is to fully leverage the potential of LLMs in creating more adaptive, intuitive, and efficient software systems.

% -------------

%
% ---- Bibliography ----
%
% \begin{thebibliography}{6}

\bibliographystyle{unsrt}

\bibliography{references} % without the .bib extension

%

% \end{thebibliography}
\end{document}